\documentclass[aps,preprint]{revtex4}%
\usepackage{amsfonts}
\usepackage{amsmath}
\usepackage{amssymb}
\usepackage{graphicx}%
\providecommand{\U}[1]{\protect\rule{.1in}{.1in}}

\begin{document}
\preprint{ }
\title[ ]{Deformed Heisenberg Algebra with a minimal length: Application to some
molecular potentials}
\author{Djamil Bouaziz}
\email{djamilbouaziz@univ-jijel.dz}
\affiliation{D\'{e}partement de Physique, Universit\'{e} de Jijel, BP 98, Ouled Aissa,
18000 Jijel, Algeria}
\begin{abstract}
We review the essentials of the formalism of quantum mechanics based on a
deformed Heisenbeg algebra, leading to the existence of a minimal length
scale. We compute in this context, the energy spectra of the pseudoharmonic
oscillator and Kratzer potentials by using a perturbative approach. We derive
the molecular constants, which characterize the vibration--rotation energy
levels of diatomic molecules, and investigate the effect of the minimal length
on each of these parameters for both potentials. We confront our result to
experimental data for the hydrogen molecule to estimate an order of magnitude
of this fundamental scale in molecular physics.

\end{abstract}

\volumeyear{2016}
\startpage{1}
\maketitle

\section{Introduction}

This work, is a continuation of the recent studies \cite{bouk,bo}, where the
effects of the minimal length on the vibration-rotation of diatomic molecules
have been investigated through the pseudoharmonic oscillator (PHO) \cite{bouk}
and the Kratzer \cite{bo} interactions.

As it is well known, the minimal length is a prediction of quantum gravity
\cite{alden,garay,rov} and string theory \cite{21,amati,konishi}, where the
Planck length is supposed to be a lower bound to all physical length scales
\cite{pad,jaekel}. This fundamental scale is implemented by the modification
of the Heisenberg uncertainty relation to the form: $\left(  \Delta X\right)
\left(  \Delta P\right)  \geq\frac{\hbar}{2}(1+\beta\left(  \Delta P\right)
^{2}+...)$, where $\beta$ is a small positive parameter
\cite{garay,amati,magiore}. The minimal uncertainty in position (minimal
length) implied by this generalized uncertainty principle (GUP) is then given
by $\left(  \Delta X\right)  _{\min}=\hbar\sqrt{\beta}$ \cite{k1}.

The study of theoretical and physical implications of the GUP is still a hot
topic. Many problems with great physical interest have been studied in
connection with the GUP, see, for instance, Refs.
\cite{k1,k2,k3,k4,brau,chang,sandor,stetsko,boua,boua3,brau2,dirac,cha,do,casimir,black-body,therm,N,unr,hossenfelder}%
.

It has been in particular concluded that the presence of the minimal length in
the formalism provides a natural ultraviolet regularization in quantum
mechanics \cite{boua} and in quantum field theory \cite{k4}. Furthermore,
upper bounds for this elementary length have been estimated; the values differ
from one application to another and, mostly, belong to the range 10$^{-6}-$
10$^{6}$ fm \cite{boua3,brau2}. The minimal length seems to depend on the
energy scale of the problem and might therefore characterize the size of the
system under study \cite{k2,boua}. The latter finding was behind the
motivation of our recent investigations \cite{bouk,bo} on the GUP effects in
diatomic molecules, because the spatial extension of these systems is
relatively large, and the effect of the minimal length may clearly manifest.

In Ref. \cite{bouk}, we studied the vibration-rotation energy levels of
diatomic molecules in the presence of a minimal length by addressing the
Schr\"{o}dinger equation with the PHO potential. A more detailed investigation
was carried out in Ref. \cite{bo}, by taking the Kratzer interaction. It has
been explicitly shown that the minimal length would have some physical
importance in studying the spectra of diatomic molecules. Furthermore, an
upper bound for the minimal length has been estimated of about $0.01$ \AA \ by
confronting the correction of the GUP to an experimental result of the
hydrogen molecule \cite{bo}.

Here, we review the main results obtained in \cite{bouk,bo}: We give the
expression of the vibration-rotation energy spectrum of diatomic molecules in
the presence of a minimal length by studying the deformed Schr\"{o}dinger
equation with the two potentials. In both cases, we apply the obtained
formulas to compute the spectroscopic constants of diatomic molecules and
investigate the effect of the GUP on these constants. We show how the
importance that the deformation parameter can play in fitting the experimental
results. Moreover, the expressions of the molecular constants derived from the
energy spectrum of the PHO show the importance of the generalization of these
studies to the case of a two-parameter deformed Heisenberg algebra, which is
presented in Sec. II.

The rest of this paper is organized as follows. In Sec. II, we review the
essentials of the formalism of quantum mechanics with a GUP. Sec. III is
devoted to investigate the KP in this formalism: we derive the molecular
constants, and investigate the effect of the minimal length on
vibration-rotation energy levels of diatomic molecules. In Sec. VI, we perform
the same study with the PHO interaction. The last section summarizes our main
results and conclusions.

\section{Deformed Heisenberg algebra with a minimal length: A review}

As mentioned in Sec. I, to include a minimal length scale, the uncertainty
relation can be modified to the form \cite{konishi,magiore} $\left(  \Delta
X\right)  \left(  \Delta P\right)  \geq\frac{\hbar}{2}(1+\beta\left(  \Delta
P\right)  ^{2}+...)$, where $\beta$ is a small positive parameter related to
the minimal length by $\left(  \Delta X\right)  _{\min}=\hbar\sqrt{\beta}$.
This generalized uncertainty principle (GUP) implies the modification of the
commutation relations between the position and the momentum operators as
\cite{k1}
\begin{equation}
\lbrack\widehat{X},\widehat{P}]=i\hbar(1+\beta\widehat{P}^{2}).
\end{equation}
The operators $\widehat{X}$ and $\widehat{P}$ can be represented in both
coordinate and momentum spaces as follow:

in momentum space, the simplest realization is \cite{k1}
\begin{equation}
\widehat{X}=(1+\beta\widehat{p}^{2})\widehat{x},\ \ \ \ \ \ \ \widehat
{P}=\widehat{p},
\end{equation}
where $\widehat{p}=p$ and$\ \widehat{x}=i\hbar\tfrac{\partial}{\partial p}$
represent the position and momentum operators of ordinary quantum mechanics.

In coordinate space one has \cite{N}
\begin{equation}
\widehat{X}=\widehat{x},\text{ \ \ \ \ \ \ \ }\widehat{P}=\text{\ }%
(1+\tfrac{1}{3}\beta\widehat{p}^{2})\widehat{p},
\end{equation}
where $\widehat{x}=x,\ \ \widehat{p}=-i\hbar\tfrac{\partial}{\partial x}$.

\bigskip The formalism of of quantum mechanics with a minimal length has been
extended to arbitrary dimensions ($D$) \cite{k1,k2,k3,k4}. The modified
Heisenberg algebra reads%
\begin{align}
\lbrack\widehat{X}_{i},\widehat{P}_{j}]  &  =i\hbar\lbrack(1+\beta\widehat
{P}^{2})\delta_{ij}+\beta^{^{\prime}}\widehat{P}_{i}\widehat{P}_{j}],\text{
\ \ }(\beta,\beta^{^{\prime}})>0,\nonumber\\
\lbrack\widehat{P}_{i},\widehat{P}_{j}]  &  =0,\label{1}\\
\lbrack\widehat{X}_{i},\widehat{X}_{j}]  &  =i\hbar\frac{2\beta-\beta
^{^{\prime}}+\beta(2\beta+\beta^{^{\prime}})\widehat{P}^{2}}{1+\beta
\widehat{P}^{2}}(\widehat{P}_{i}\widehat{X}_{j}-\widehat{P}_{j}\widehat{X}%
_{i}).\nonumber
\end{align}

The corresponding GUP is
\begin{equation}
\left(  \Delta X_{i}\right)  \left(  \Delta P_{i}\right)  \geq\tfrac{\hbar}%
{2}(1+\beta\sum\limits_{j=1}^{N}[\left(  \Delta P_{j}\right)  ^{2}%
+\left\langle \widehat{P}_{j}\right\rangle ^{2}]+\beta^{^{\prime}}[\left(
\Delta P_{i}\right)  ^{2}+\left\langle \widehat{P}_{i}\right\rangle ^{2}]),
\end{equation}
in which the minimal uncertainty in position (minimal length) is \cite{k2}%
\begin{equation}
\left(  \Delta X_{i}\right)  _{\min}=\hbar\sqrt{D\beta+\beta^{^{\prime}}%
},\text{ \ \ }\forall i.
\end{equation}

The position and momentum operators are represented in momentum space as
\cite{k2,chang}%
\begin{equation}
\widehat{X}_{i}=i\hbar\lbrack\left(  1+\beta p^{2}\right)  \tfrac{\partial
}{\partial p_{i}}+\beta^{\prime}p_{i}p_{j}\tfrac{\partial}{\partial p_{j}%
}+\gamma p_{i}],\text{ \ \ \ \ \ }\widehat{P}_{i}=p_{i}, \label{re}%
\end{equation}
where $\gamma$ is a small positive parameter related to $\beta$ and
$\beta^{\prime}$.

In coordinate space, the simplest representation of the operators $\widehat
{X}_{i}$ and $\widehat{P}_{i}$ is \cite{brau}%
\begin{equation}
\widehat{X}_{i}=\widehat{x}_{i},\text{ \ \ \ }\widehat{P}_{i}=\widehat{p}%
_{i}\left(  1+\beta\widehat{p}^{2}\right)  , \label{b}%
\end{equation}
where $\widehat{x}_{i}$ and $\widehat{p}_{i}$ satisfy the standard commutation
relations of ordinary quantum mechanics.

Representation (\ref{b}) satisfies the deformed algebra in the case
$\beta^{\prime}=2\beta$ up to the first order of $\beta$. It is especially
more adequate in the treatment of the minimal length as a perturbation to the
Schr\"{o}dinger equation for a given interaction. In this special case the
minimal length reads in 3-dimensions as
\begin{equation}
\left(  \Delta X_{i}\right)  _{\min}=\hbar\sqrt{5\beta},\text{ \ \ }\forall i.
\end{equation}

The Schr\"{o}dinger equation with the representation (\ref{brau}) can be
written as follow:%
\begin{equation}
\left(  \frac{\widehat{p}^{2}}{2\mu}+V(r)+\frac{\beta}{\mu}\widehat{P}%
^{4}\right)  \psi(\mathbf{r})=E\psi(\mathbf{r}), \label{6}%
\end{equation}
where terms of order $\beta^{2}$ have been neglected.

In the following sections, we will use equation (\ref{6}) to study the Kratzer
and the PHO potentials. The main goal of this study is to derive the
expressions of the spectroscopic constants of diatomic molecules in the
presence of a minimal length by using the deformed energy spectrum of these
two interactions.

\section{Energy spectrum of Kratzer potential with a minimal length}

We are interested to the deformed Sch\"{o}dinger equation (\ref{6}) with the
Kratzer's molecular potential (KP), which has the form \cite{flug}%
\begin{equation}
V(r)=g_{1}/r^{2}-g_{2}/r, \label{k}%
\end{equation}
with $g_{1}=D_{e}r_{e}^{2}$ and $g_{2}=2D_{e}r_{e}$, where $D_{e}$ is the
dissociation energy and $r_{e}$ is the equilibrium internuclear distance of a
given diatomic molecule.

It has been shown in detail in Ref. \cite{bo} that the minimal length
corrections $\Delta E_{n\ell}$ can be analytically computed, and the energy
spectrum of the KP in the presence of a minimal length is given by the
following expression:%
\begin{align}
E_{n\ell}  &  =-\tfrac{\gamma^{2}D_{e}}{\left(  \lambda+n\right)  ^{2}}%
+\beta\mu D_{e}^{2}\left(  \tfrac{2\gamma}{\lambda+n}\right)  ^{4}\left\{
-\tfrac{3}{4}+\tfrac{\lambda+n}{\lambda-1/2}\left(  1+\tfrac{\gamma^{2}}%
{2}\left(  \tfrac{1}{\left(  \lambda+n\right)  ^{2}}-\tfrac{2}{\lambda
(\lambda-1)}\right)  \right)  \right. \nonumber\\
&  \left.  +\tfrac{\gamma^{4}}{4}\tfrac{1}{(\lambda-\frac{1}{2})(\lambda
-1)(\lambda-\frac{3}{2})(\lambda+n)}\left(  1+\tfrac{3n(2\lambda+n)}%
{\lambda(2\lambda+1)}\right)  \right\}  , \label{sss}%
\end{align}
with the notations%
\begin{equation}
\text{\ }\gamma=\tfrac{r_{e}}{\hbar}\sqrt{2\mu D_{e}},\text{ \ \ }%
\lambda=1/2+\sqrt{(\ell+1/2)^{2}+\gamma^{2}},
\end{equation}
where $n$ and $\ell$ are, respectively, the radial (vibrational) and orbital
(rotational) quantum numbers; \ and $\mu\ $is the reduced mass of the molecule.

Formula (\ref{sss}) shows the effect of this deformed algebra on the energy
levels of KP. As it has been outlined in Ref. \cite{bo}, several features of
diatomic molecules can be studied by using this formula. In the following, we
apply Eq. (\ref{sss}) to investigate the effect of the minimal length on the
spectroscopic constants, which characterize the rovibrational energy levels of
diatomic molecules.

\begin{quote}
\textbf{Spectroscopic constants of diatomic molecules}
\end{quote}

As we will see, the energy spectrum of the KP is similar to the well-known
spectroscopic formula which defines the vibration-rotation energy spectrum of
diatomic molecules \cite{ZPE,flug}.

To this end, we observe that the dimensionless parameter $\gamma$, in Eq.
(\ref{sss}), is so large for most molecules ($\gamma\gg1$) \cite{flug}, the
levels $E_{n\ell}$ may then be expanded into powers of $1/\gamma$ as follow:%
\begin{align}
E_{n\ell}  &  =D_{e}\left(  -1+2(n+\tfrac{1}{2})\tfrac{1}{\gamma}+(\ell
+\tfrac{1}{2})^{2}\tfrac{1}{\gamma^{2}}\allowbreak-3(n+\tfrac{1}{2})^{2}%
\tfrac{1}{\gamma^{2}}+4(n+\tfrac{1}{2})^{3}\tfrac{1}{\gamma^{3}}-3(n+\tfrac
{1}{2})(\ell+\tfrac{1}{2})^{2}\tfrac{1}{\gamma^{3}}\right) \nonumber\\
&  +\beta\mu D_{e}^{2}\left(  6\left\{  (n+\tfrac{1}{2})^{2}+\tfrac{1}%
{4}\right\}  \tfrac{1}{\gamma^{2}}\allowbreak+2(n+\tfrac{1}{2})\left\{
-\tfrac{1}{4}+4(\ell+\tfrac{1}{2})^{2}-15(n+\tfrac{1}{2})^{2}\right\}
\tfrac{1}{\gamma^{3}}\right)  +... \label{r}%
\end{align}

This formula shows the different parts of the undeformed spectrum ($\beta=0$),
and the corrections caused by the minimal length. The effect of this GUP on
each kind of the rovibrational energy was discussed in detail in Ref.
\cite{bo}. Here, we will suggest some ideas on how to get information about
the order of the deformation parameter $\beta$ from formula (\ref{r}), and
then apply it to extract the spectroscopic constants of diatomic molecules,
which is a way to give values of $\beta$ for any molecule.

We now have basically two approaches to deal with the parameter $\beta$ in
formula (\ref{r}): The first is to consider that $\beta$ is independent of the
molecular constants $r_{e}$ and $D_{e}$, which have known values in molecular
spectroscopy. Therefore, an upper bound for $\beta$ can be estimated by
assuming, e.g, that the minimal length corrections in in Eq. (\ref{r}) are
included in the gap between the experimental results and those predicted by
the formula (\ref{r}) in the case case $\beta=0$. It has been argued in Ref.
\cite{bo} that a better estimate may be obtained by considering the
vibrational ground-state energy $E_{00}$ of the hydrogen molecule (H$_{2}$).
This led to an upper bound for the minimal length with the value
\begin{equation}
\left(  \Delta X\right)  _{\min}=\hbar\sqrt{5\beta}<0.01\text{ \AA .}%
\end{equation}
The second view point consists of looking at formula (\ref{r}) as an energy
spectrum of a three-parameter potential, i.e., $D_{e},$ $r_{e},$ and $\beta$;
so, the values of $\beta$ depend now on $D_{e}$ and $r_{e}$. This viewpoint
allows for adjusting the three parameters of the "deformed KP" with the
spectroscopic data. To this end, we can suggest to extract, from the deformed
energy spectrum (\ref{r}), the spectroscopic constants for diatomic molecules.

The commonly used energy equation in molecular spectroscopy is \cite{ZPE,spec}%
\begin{align}
E_{n\ell}  &  =Y_{00}+\omega_{e}\left(  n+\tfrac{1}{2}\right)  -\omega
_{e}x_{e}(n+\tfrac{1}{2})^{2}+\omega_{e}y_{e}(n+\tfrac{1}{2})^{3}\nonumber\\
&  +B_{e}\ell(\ell+1)-\alpha_{e}\left(  n+\tfrac{1}{2}\right)  \ell(\ell+1),
\label{sp}%
\end{align}
where%
\begin{equation}
Y_{00}=\tfrac{D_{e}}{4}\tfrac{\omega_{e}x_{e}}{\gamma^{2}}+\tfrac{\alpha
_{e}\omega_{e}}{12B_{e}}(1+\tfrac{\omega_{e}\omega_{e}}{12B_{e}^{2}}).
\end{equation}

The values of the spectroscopic constants $\omega_{e},$ $\omega_{e}x_{e},$
$\omega_{e}y_{e},$ $B_{e}$ and $\alpha_{e}$ are determined by fitting the
spectroscopic data for any molecules \cite{ZPE}.

We write now equation (\ref{r}) in the form
\begin{align}
E_{n\ell}  &  =D_{e}\left[  -1+2(n+\tfrac{1}{2})\tfrac{1}{\gamma}+(\ell
+\tfrac{1}{2})^{2}\tfrac{1}{\gamma^{2}}\allowbreak-3(n+\tfrac{1}{2})^{2}%
\tfrac{1}{\gamma^{2}}\right. \nonumber\\
&  \left.  +4(n+\tfrac{1}{2})^{3}\tfrac{1}{\gamma^{3}}-3(n+\tfrac{1}{2}%
)(\ell+\tfrac{1}{2})^{2}\tfrac{1}{\gamma^{3}}\right] \nonumber\\
&  +\beta\mu D_{e}^{2}\left[  6\left\{  (n+\tfrac{1}{2})^{2}+\tfrac{1}%
{4}\right\}  \tfrac{1}{\gamma^{2}}\allowbreak-\tfrac{1}{2}(n+\tfrac{1}%
{2})\right. \nonumber\\
&  +\left.  2(n+\tfrac{1}{2})\left\{  4(\ell+\tfrac{1}{2})^{2}-15(n+\tfrac
{1}{2})^{2}\right\}  \tfrac{1}{\gamma^{3}}\right]  +...\text{ \ .} \label{S}%
\end{align}
The identification of Eqs. (\ref{S}) and (\ref{sp}) leads to the following
deformed spectroscopic constants:%
\begin{align}
Y_{00}  &  =\tfrac{1}{4}\tfrac{D_{e}}{\gamma^{2}}+\tfrac{3}{2}\beta\mu
\tfrac{D_{e}^{2}}{\gamma^{2}},\text{ \ }\omega_{e}=\tfrac{2D_{e}}{\gamma
}-\tfrac{3}{4}\tfrac{D_{e}}{\gamma^{3}}+\tfrac{3}{2}\beta\mu\tfrac{D_{e}^{2}%
}{\gamma^{3}},\nonumber\\
\omega_{e}x_{e}  &  =\tfrac{3D_{e}}{\gamma^{2}}-6\beta\mu\tfrac{D_{e}^{2}%
}{\gamma^{2}},\text{ \ }\omega_{e}y_{e}=\tfrac{4D_{e}}{\gamma^{3}}-30\beta
\mu\tfrac{D_{e}^{2}}{\gamma^{3}},\label{s}\\
B_{e}  &  =\tfrac{D_{e}}{\gamma^{2}},\text{ \ \ }\alpha_{e}=\tfrac{3D_{e}%
}{\gamma^{3}}-8\beta\mu\tfrac{D_{e}^{2}}{\gamma^{3}},\nonumber
\end{align}

We observe that the leading corrections of the minimal length are of order
$1/\gamma^{2}$ and concerns the constant of anharmonicity of vibrations
$\omega_{e}x_{e}$ and the constant $Y_{00}$, which does not influence the line
positions in a spectrum. However, the rotational constant $B_{e}$ is not
affected by this deformed algebra.

As the experimental values of the spectroscopic constants are available for
diatomic molecules \cite{ZPE}, it follows that the expressions (\ref{s}) can
be used to give values of $\beta$ for each molecule. This investigation is
under consideration in the general case of deformed Heisenberg algebra
(\ref{1}), with $\beta^{\prime}\neq2\beta$.

\section{Energy spectrum of the pseudoharmonic oscillator with a minimal
length}

The pseudoharmonic oscillator (PHO) is also one of the molecular interactions,
which is used in the study of the vibration-rotation\ spectra of diatomic
molecules. The form of this potential is
\begin{equation}
V(r)=D_{e}(\frac{r}{r_{e}}-\frac{r_{e}}{r})^{2},
\end{equation}
where $D_{e}$ is the dissociation energy and $r_{e}$ is the equilibrium
internuclear distance of a given diatomic molecule.

It has been shown in Ref. \cite{bouk} that the minimal length correction of
this potential can also be analytically derived by following the same
procedures as in Sec. III. The deformed energy spectrum of the PHO reads
\cite{bouk}\
\begin{align}
E_{n\ell}  &  =E_{n\ell}^{0}+4\mu\beta\left(  (E_{n\ell}^{0})^{2}%
+4D_{e}E_{n\ell}^{0}+6D_{e}^{2}-(4D_{e}^{2}+2D_{e}E_{n\ell}^{0})\frac
{\lambda+2n+1}{\gamma}+\right. \nonumber\\
&  \left.  D_{e}^{2}\frac{\lambda^{2}+(6n+3)\lambda+6n(n+1)+2}{\gamma^{2}%
}-\gamma\frac{2D_{e}\left(  2D_{e}+E_{n\ell}^{0}\right)  }{\lambda}+D_{e}%
^{2}\gamma^{2}\frac{\lambda+2n+1}{\lambda(\lambda^{2}-1)}\right)  ,
\label{E12}%
\end{align}
where $E_{n\ell}^{0}$ is the undeformed spectrum, given by
\begin{equation}
E_{n\ell}^{0}=-2D_{e}\left(  1-\tfrac{1}{\gamma}\left(  2n+1+\lambda\right)
\right)  , \label{E4}%
\end{equation}
with the notations
\begin{equation}
\text{ }\lambda=\sqrt{\gamma^{2}+(\ell+\tfrac{1}{2})^{2}},\text{ \ }%
\gamma=\ \sqrt{\tfrac{2\mu D_{e}r_{e}^{2}}{\hbar^{2}}},\text{ \ }%
\end{equation}
where, $n=0,1,2,\ldots$, and $\ell=0,1,2,\ldots$are, respectively, the radial
(vibrational) and orbital (rotational) quantum numbers, and $\mu\ $is the
reduced mass of the molecule.

The effect of the GUP on the vibration-rotation energy levels of a given
diatomic molecule with the PHO interaction was qualitatively investigated in
Ref. \cite{bouk} by using formula (\ref{E12}). Here, we use this deformed
spectrum to give the expressions of the molecular constants in the presence of
a minimal length.

\textbf{Spectroscopic constants of diatomic molecules}

Formula (\ref{E12}) might be viewed as an energy spectrum of a three-parameter
potential function, i.e., $D_{e},$ $r_{e},$ and $\beta$. One can then derive
the molecular constants for diatomic molecule by expanding $E_{n\ell}$ in Eq.
(\ref{E12}) into powers of $1/\gamma$ ($\gamma\gg1$); this leads to the
expression%
\begin{align}
E_{n,\ell}  &  =(\tfrac{1}{4}+6\mu\beta D_{e})\tfrac{D_{e}}{\gamma^{2}}%
+\tfrac{4D_{e}}{\gamma}\left(  1+\tfrac{3\mu\beta D_{e}^{2}}{\gamma^{2}%
}\right)  (n+\tfrac{1}{2})+\tfrac{24\mu\beta D_{e}^{2}}{\gamma^{2}}%
(n+\tfrac{1}{2})^{2}\nonumber\\
&  +\tfrac{D_{e}}{\gamma^{2}}\ell(\ell+1)+\tfrac{16\mu\beta D_{e}^{2}}%
{\gamma^{3}}\left(  n+\tfrac{1}{2}\right)  \ell(\ell+1)+....\text{ ,}%
\end{align}
which is of the form of the spectroscopic formula (\ref{sp}), with the
following spectroscopic constants:%
\begin{align}
Y_{00}  &  =(\tfrac{1}{4}+6\mu\beta D_{e})\tfrac{D_{e}}{\gamma^{2}},\text{
\ \ }\omega_{e}=\tfrac{4D_{e}}{\gamma}\left(  1+\tfrac{3\mu\beta D_{e}^{2}%
}{\gamma^{2}}\right) \nonumber\\
\omega_{e}x_{e}  &  =-\tfrac{24\mu\beta D_{e}^{2}}{\gamma^{2}},\text{
\ \ }B_{e}=\tfrac{D_{e}}{\gamma^{2}},\text{ \ \ \ \ \ \ \ }\alpha_{e}%
=-\tfrac{16\mu\beta D_{e}^{2}}{\gamma^{3}}.
\end{align}
The value of the constant $\omega_{e}y_{e}$ is zero in the order $1/\gamma
^{3}.$

In contrast to ordinary quantum mechanics ($\beta=0$), where the spectrum of
the PHO leads to zero values of the spectroscopic constants $\omega_{e}x_{e}$
and $\alpha_{e}$, in the presence of a minimal length, these constants depend
on $\beta$ and have nonzero values. The basic empirical terms, known in the
vibration-rotation energy spectrum of diatomic molecules, are now present in
this deformed version of quantum mechanics. However, the sign of $\omega
_{e}x_{e}$ and $\alpha_{e}$ is conflicting compared to that of the
experimental values, at least for the diatomic molecules listed in Ref.
\cite{ZPE}.

We can conclude that the PHO in this one-parameter deformed algebra could not
be used to fit the deformed spectroscopic constants with the empirical
results. It follows that the extension of this work to the general case
$\beta^{\prime}\neq2\beta$\ is interesting; it allows us not only to adjust
the parameters $D_{e},$ $r_{e},$ $\beta$ and $\beta^{\prime}$ but also to
establish a constraint between the deformation parameters $\beta$ and
$\beta^{\prime}$ by appropriately choosing the sign of the molecular
constants. This study is in finalization and would be published else where.

\section{Summary and conclusions}

We have investigated, in quantum mechanics with a deformed Heisenberg algebra
including a minimal length $\left(  \Delta X_{i}\right)  _{\min}=\hbar
\sqrt{5\beta}$, the vibration-rotation energy spectra of diatomic molecules by
two molecular interactions: the Kratzer and the PHO potentials. We have
discussed how an order of the deformation parameter $\beta$ can be evaluated
from the minimal length corrections and the spectroscopic data of diatomic
molecules. With the Kratzer's potential, an upper bound of the minimal length
of about 0.01 \AA \ has been obtained by comparing the theoretical and
experimental values of the vibrational ground-state energy of the molecule
H$_{2}$. On the other hand, by supposing that the deformation parameter is a
third parameter of the potentials, we have derived the spectroscopic constants
of diatomic molecules with both interactions and we have examined the effects
of the minimal length on each of these constants. We showed that in the case
of the PHO the extension of this study to the general case of a two-parameter
deformed Heisenberg algebra would be mandatory and interesting because it
allows to give some physical constraint between $\beta$ and $\beta^{\prime}$.

\begin{acknowledgments}
This work is supported by the Algerian Ministry of Higher Education and
Scientific Research, under the CNEPRU Project No. D01720140007.
\end{acknowledgments}

\end{document}